\documentclass[
    reprint, twocolumn,
    aps,prl,10pt,amsmath,amssymb,
    longbibliography,superscriptaddress,
    showpacs,preprintnumbers,
]{revtex4-2}

\usepackage{xr-hyper}
\usepackage[colorlinks=true, linkcolor=blue, citecolor=blue, urlcolor=blue, filecolor=blue]{hyperref}
\usepackage{bm, bbm, bbold}
\usepackage{braket}
\usepackage{mathtools}
\usepackage{amsmath,amssymb,amsfonts}
\usepackage[dvipsnames]{xcolor}
\usepackage{graphicx}
\usepackage[section]{placeins}
\usepackage{multirow}
\usepackage{orcidlink}

\def\ShowRevision{0}

\ifx\ShowRevision\undefined
    
    \newcommand{\MYsout}[2]{}
    \newcommand{\MYcancel}[2]{}
    \newcommand{\MYcomment}[3]{}
\else
    \newcommand{\MYsout}[2]{\color{#1}{\sout{#2}}}  
    \newcommand{\MYcancel}[2]{\color{#1}{\cancel{#2}}}  
    \newcommand{\MYcomment}[3]{\color{#1}{[\small \textbf{{#2}:} \textit{#3}]}}  
\fi

\newcommand{\mb}[1]{{\bm{\mathrm{#1}}}}  

\newcommand{\veps}[0]{\varepsilon}

\renewcommand{\Re}{\operatorname{Re}}
\renewcommand{\Im}{\operatorname{Im}}
\newcommand{\dd}[0]{{\mathrm{d}}}
\newcommand{\tbar}[0]{{\text{-}}}  

\newcommand{\Eq}[1]{Eq.~\eqref{#1}}
\newcommand{\Eqs}[1]{Eqs.~\eqref{#1}}
\newcommand{\Equ}[1]{Equation~\eqref{#1}}

\newcommand{\Sec}[1]{Sec.~\ref{#1}}

\newcommand{\Fig}[1]{Fig.~\ref{#1}}

\newcommand{\Figu}[1]{Figure~\ref{#1}}

\newcommand{\bq}[0]{{\mb{q}}}

\newcommand{\nk}[0]{{n\mb{k}}}

\newcommand{\mk}[0]{{m\mb{k}}}
\newcommand{\mnk}[0]{{mn\mb{k}}}
\newcommand{\mkq}[0]{{m\mb{k+q}}}

\newcommand{\qnu}[0]{{\mb{q}\nu}}

\newcommand{\nint}{\!\int\!} 
\newcommand{\nbint}[2]{\!\int_{#1}^{#2}\!\!} 

\DeclarePairedDelimiterX\mel[3]{\langle}{\rangle}{#1 \delimsize\vert\mathopen{} #2 \delimsize\vert\mathopen{} #3}
\DeclarePairedDelimiterX\abs[1]{\lvert}{\rvert}{#1}

\definecolor{DARKGREEN}{rgb}{0,0.5,0}

\newcommand{\EPjl}[0]{\texttt{ElectronPhonon.jl}}

\makeatletter
\newcommand*{\addFileDependency}[1]{
  \typeout{(#1)}
  \@addtofilelist{#1}
  \IfFileExists{#1}{}{\typeout{No file #1.}}
}
\makeatother

\newcommand*{\myexternaldocument}[2]{
    \externaldocument[#2]{#1}
    \addFileDependency{#1.tex}
    \addFileDependency{#1.aux}
}

\myexternaldocument{supp}{supp-}

\begin{document}

\title{Nonperturbative self-consistent electron-phonon spectral functions and transport}

\author{Jae-Mo Lihm\,\orcidlink{0000-0003-0900-0405}}
\email{jaemo.lihm@gmail.com}
\affiliation{%
European Theoretical Spectroscopy Facility, Institute of Condensed Matter and Nanosciences, Universit\'e catholique de Louvain, Chemin des \'Etoiles 8, B-1348 Louvain-la-Neuve, Belgium
}%
\author{Samuel Ponc\'e\,\orcidlink{0000-0003-1159-8389}}
\email{samuel.ponce@uclouvain.be}
\affiliation{%
European Theoretical Spectroscopy Facility, Institute of Condensed Matter and Nanosciences, Universit\'e catholique de Louvain, Chemin des \'Etoiles 8, B-1348 Louvain-la-Neuve, Belgium
}%
\affiliation{%
WEL Research Institute, avenue Pasteur, 6, 1300 Wavre, Belgium.
}%
\date{\today}

\begin{abstract}
Electron-phonon coupling often dominates the electron spectral functions and carrier transport properties.
However, studies of this effect in real materials have largely relied on perturbative one-shot methods due to the lack of a first-principles theoretical and computational framework.
Here, we present a self-consistent theory and implementation for the nonperturbative calculations of spectral functions and conductivity due to electron-phonon coupling.
Applying this method to monolayer InSe, we demonstrate that self-consistency qualitatively affects the spectral function and transport properties compared to state-of-the-art one-shot calculations and allow one to reconcile calculations with angle-resolved photoemission experiments.
The developed method can be widely applied to materials with dominant electron-phonon coupling at moderate computational cost.
\end{abstract}

\maketitle

\footnotetext[1]{See Supplemental Material [URL will be inserted by publisher], which includes Refs.~\cite{Dunn1975, Nery2022, Hamann2013ONCVPSP, Perdew1996, Setten2018, Sohier2017, Royo2019, Gonze2020Abinit, Giustino2007EPW, Marzari1997, Verdi2015, Sjakste2015, Brunin2020PRL, Brunin2020PRB, Jhalani2020, Park2020, Ponce2023PRL, Royo2021, Bezanson2017Julia}, for the derivation of the formula, computational details, and additional results.}
\newcommand{\citeSupp}[0]{Note1}

Spectral functions are essential in many-body physics, linking theory and experiment.
Experimentally, spectral functions can be extracted using high-resolution angle-resolved photoemission spectroscopy (ARPES)~\cite{Damascelli2003}.
Theoretically, spectral functions can be calculated from Green's functions, including many-body correlation via the self-energy.
The coupling between electrons and phonons, the collective vibration of the lattice, plays a crucial and often dominant role in shaping the electron spectral function~\cite{Giustino2017}.
The electron-phonon ($e$-ph) interaction leads to the broadening and renormalization of the quasiparticle peak~\cite{Allen1983, Allen1981, Allen1976, Marini2008, Giustino2010}, photoemission kinks in metals~\cite{Valla1999, Eiguren2003}, and additional structures such as satellite peaks~\cite{Engelsberg1963, Moser2013, Chen2015, Cancellieri2016, Wang2016, Verdi2017}.
Understanding and predicting these effects is the foundation of electron-phonon physics.

\textit{Ab initio} calculation of the $e$-ph self-energy and spectral functions is usually limited to perturbation theory.
Standard approaches use the perturbative Fan--Migdal (FM) self-energy~\cite{Migdal1958, Fan1951}, solving the Dyson equation~\cite{Engelsberg1963} or applying the cumulant expansion~\cite{Gunnarsson1994Cumulant, Aryasetiawan1996Cumulant, Guzzo2011Cumulant, Story2014Cumulant, Gumhalter2016Cumulant, Verdi2017, Nery2018, Zhou2019STO, DeAbreu2022, Kandolf2022} for spectral functions.
However, these approaches often fail, causing divergent quasiparticle dispersion~\cite{Larsen1966, Ponce2015, Kandolf2022} or unphysical spectral properties~\cite{Kandolf2022}.
Notably, this problem is not limited to strong \textit{e}-ph coupling: the negative curvature in the quasiparticle dispersion of the Fr\"ohlich model~\cite{Larsen1966} and the divergent linewidth of piezoelectric materials~\cite{Lihm2024Piezo} illustrate failures at all coupling strengths.
Advanced methods incorporating higher-order $e$-ph interactions~\cite{Prokofev1998DMC, Jeckelmann1998, Berciu2006, Bonca2019, Carbone2021, mahajan2024, Sakkinen2015A, Sakkinen2015B, Rademaker2016SCBA, Chen2016SCBA, Esterlis2018SCBA, Dee2019SCBA, Dee2020SCBA, Mitric2022, Mitric2023} are limited to model Hamiltonians.
Nonperturbative methods using molecular dynamics~\cite{Zacharias2020, Quan2024} or special-displacement supercells~\cite{Zacharias2016ZG, Zacharias2020ZG} are difficult to converge~\cite{Quan2024, Ponce2024}, and cannot capture nonadiabatic $e$-ph coupling, crucial in polar materials~\cite{Ponce2015, Miglio2020ZPR}.

The electron spectral function also governs charge transport.
Within the Green--Kubo theory~\cite{Green1954, Kubo1957}, neglecting current vertex corrections allows direct computation of conductivity from spectral functions~\cite{MahanBook, Basov2011RMP}.
This approach, known as the bubble approximation, has the advantage of including broadening and satellites beyond the quasiparticle approximation~\cite{Zhou2019STO, Chang2022, Abramovitch2023, Abramovitch2024, Quan2024}.

A material of particular interest for nonperturbative $e$-ph coupling is monolayer InSe.
While known for high electron mobility~\cite{Bandurin2017InSe, Wan2023InSe}, its valence band features a flat, Mexican-hat-like dispersion~\cite{Zolyomi2014InSe, Rybkovskiy2014InSe, Debbichi2015InSe, Magorrian2016InSe, Kibirev2018InSe, Graham2021InSe, Stepanov2022InSe}, leading to strong $e$-ph interaction.
In particular, state-of-the-art one-shot calculations predicted a dynamical splitting of the valence band into two peaks in the spectral function with a gap larger than 150~meV~\cite{Lugovskoi2019InSe}.
The hole mobility of monolayer InSe has been studied using the Boltzmann transport equation (BTE)~\cite{Das2019InSe, Li2019InSe, Ponce2023PRB} and the bubble approximation with the one-shot spectral function~\cite{Lugovskoi2019InSe}.
Yet, given the strong renormalization, nonperturbative corrections are expected to be essential in monolayer InSe.

In this Letter, we present a nonperturbative \textit{ab initio} calculation of the $e$-ph spectral function and conductivity.
Our approach, the self-consistent $GD_0$ (sc$GD_0$) method~\cite{Sakkinen2015A, Sakkinen2015B, Rademaker2016SCBA, Chen2016SCBA, Esterlis2018SCBA, Dee2019SCBA, Dee2020SCBA, Mitric2022, Mitric2023}, extends the FM self-energy to include the self-consistently renormalized spectral functions, analogously to the sc$GW_0$ method for electron-electron interactions~\cite{Shishkin2007GW}.
Using self-consistent spectral functions, we compute conductivity using the bubble approximation.
We find that self-consistency significantly affects the spectral function and conductivity of monolayer InSe.
This effect is present in any materials and can be explored with the sc$GD_0$ method.


\begin{figure*}[tb]
\centering
\includegraphics[width=0.99\linewidth]{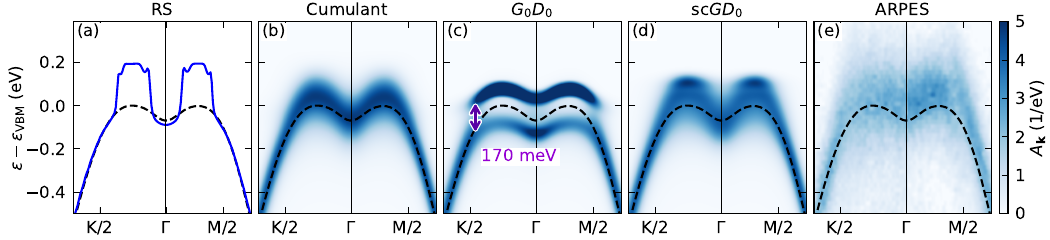}
\caption{
(a-d) Renormalized dispersion and spectral functions of monolayer InSe at $T=100$~K computed using the Rayleigh--Schr\"odinger (RS) perturbation theory~[\Eq{eq:RSPT}], cumulant approximation~[Eqs.~(\ref{eq:cum_G}, \ref{eq:cum_C})],  one-shot $G_0 D_0$~[Eqs.~(\ref{eq:G0D0_A}, \ref{eq:G0D0_G})], and sc$GD_0$~[Eqs.~(\ref{eq:GD0_Im}-\ref{eq:GD0_Re_window})] methods.
The dashed black curves represent the bare electron band.
Results in (a-c) are computed with an artificial broadening of $\eta$ = 5~meV but the results are qualitatively the same with larger values; see \Sec{supp-sec:InSe_additional} of SM~\cite{\citeSupp}.
The sc$GD_0$ result in (d) does not use any artificial broadening.
To account for the experimental resolution, we apply a Gaussian filter with widths of $0.03~\mathrm{\AA}^{-1}$ and $20~\mathrm{meV}$~\cite{Hamer2019InSe} on the calculated spectral functions in (b-d).
This filter does not qualitatively impact the results, as shown in \Fig{supp-fig:spectral_noblur} of SM~\cite{\citeSupp} for the results without the filter.
(e) Experimental ARPES photointensity reported in Ref.~\cite{Hamer2019InSe}, with the background value (average of the data inside the band gap) subtracted, and the data at each $\mb{k}$ normalized to satisfy $\nint A_\mb{k}(\omega) \dd\omega = 1$.
The data is vertically shifted to align the intensity peaks with the bare bands at $\veps=-0.4~$eV ($\veps=0$ is 2.234~eV below the experimental Fermi level).
}
\label{fig:spectral}
\end{figure*}

The most common approximation for the $e$-ph self-energy is the FM formula~\cite{Giustino2017}:
\begin{multline} \label{eq:G0D0_FM}
    \Sigma_\nk^{G_0 D_0\tbar{\rm FM}}(\veps) = \sum_{\nu m}  \nint \frac{\dd\mb{q}}{V^{\rm BZ}}  \abs{g_{mn\nu}(\mb{k},\mb{q})}^2 \\
    \times \sum_{\pm} \frac{f^\pm(\veps_\mkq) + n_\qnu}{\veps - \veps_\mkq \pm \omega_\qnu + i\eta},
\end{multline}
where $V^{\rm BZ}$ is the Brillouin zone volume,
$g_{mn\nu}(\mb{k,q})$ the $e$-ph matrix element,
$\veps_\nk$ and $\omega_\qnu$ the electron and phonon energies,
$\eta$ is a positive infinitesimal broadening,
$f^+(\veps)$ the Fermi--Dirac distribution, $f^-(\veps) = 1 - f^+(\veps)$,
and $n_\qnu$ the Bose-Einstein distribution.
We neglect the band off-diagonal self-energy, which is a valid approximation if the bands do not hybridize~\cite{Lihm2020}, as in InSe.
We refer to \Eq{eq:G0D0_FM} as the $G_0 D_0$-FM self-energy, corresponding to a one-loop diagram (see \Fig{supp-fig:feynman}(a) of Supplemental Material (SM)~\cite{\citeSupp}) with bare electron and phonon Green's functions, $G_0$ and $D_0$, as in the $G_0 W_0$ approximation~\cite{1986HybertsenGW}.
In this Letter, we assume that $G_0$ is obtained from Kohn--Sham density functional theory and that $D_0$ and $e$-ph coupling come from adiabatic density functional perturbation theory~\cite{Gonze1997, Baroni2001, Giustino2017}.
We neglect the renormalization of the phonon Green's function since we focus on the intrinsic and low doping regime where the phonon renormalization is small~\cite{Giustino2017}.
The total $G_0 D_0$ self-energy also contains the Debye--Waller (DW) term~\cite{Allen1976, Allen1981, Allen1983},
\begin{equation} \label{eq:G0D0}
    \Sigma^{G_0 D_0}_\nk(\veps)
    = \Sigma^{G_0 D_0\tbar{\rm FM}}_\nk(\veps) + \Sigma^{\rm DW}_\nk \,.
\end{equation}
In this work, we compute the DW term within the rigid-ion approximation~\cite{Ponce2014PRB}.

Using the $G_0 D_0$ self-energy, the renormalized electron dispersion can be computed within Rayleigh--Schr\"odinger (RS) perturbation theory~\cite{Gonze2011, Ponce2014, Ponce2015, Giustino2017, Nery2018} as
\begin{equation} \label{eq:RSPT}
    E^{\rm RS}_\nk
    = \veps_\nk + \Sigma^{G_0 D_0}_\nk(\veps_\nk)\,.
\end{equation}
The energy-dependent spectral function is computed by solving the Dyson equation
\begin{align}
    \label{eq:G0D0_A}
    A^{G_0 D_0}_\nk(\veps) &= -\frac{1}{\pi} \Im G^{G_0 D_0}_\nk(\veps) \,, \\
    \label{eq:G0D0_G}
    G^{G_0 D_0}_\nk(\veps) &= \frac{1}{\veps - \veps_\nk + \mu - \Sigma^{G_0 D_0}_\nk(\veps)} \,,
\end{align}
with $\mu$ the chemical potential.

To capture higher-order $e$-ph coupling, one often uses the cumulant approximation.
It replaces the Dyson equation with the expansion
\begin{equation} \label{eq:cum_G}
    G^{\rm Cum.}_\nk(\veps) =-i \nint \mathrm{d}t \, e^{i\veps t}   \Theta(t) e^{-i\veps_\nk t} e^{C_\nk(t)} \,,
\end{equation}
where $\Theta(t)$ is the Heaviside function and $C_\nk(t)$ the cumulant function.
We adopt the retarded cumulant method~\cite{Kas2014, Kas2017, Zhou2019STO} as we deal with finite temperatures.
The cumulant function is approximated via second-order perturbation theory~\cite{Kas2014, Story2014Cumulant, Kas2017},
\begin{equation} \label{eq:cum_C}
    C_\nk(t) \approx \frac{1}{\pi} \nint \mathrm{d}\veps \, \abs{\Im \Sigma_\nk^{G_0 D_0}(\veps + \veps_\nk)} \frac{e^{-i\veps t} + i\veps t - 1}{\veps^2} \,.
\end{equation}
Unlike the $G_0 D_0$ approximation, which only produces a single satellite~\cite{Verdi2017}, the cumulant approach can generate multiple satellite peaks, as seen in experiments~\cite{Moser2013, Chen2015, Cancellieri2016, Wang2016}.
However, it has been shown that the cumulant approximation can lead to unphysically large renormalizations for $\mb{k}$ points away from the band edge~\cite{Kandolf2022, Mitric2023}, similar to the RS approximation~\cite{Larsen1966}.

To incorporate nonperturbative $e$-ph effects, we adopt the sc$GD_0$ method, where the self-energy is obtained by replacing the bare electron Green's function in the $G_0 D_0$ approximation with the dressed one [see \Fig{supp-fig:feynman}(b) in SM~\cite{\citeSupp}].
The imaginary part of the sc$GD_0$ self-energy is given by~\cite{Abramovitch2023}
\begin{multline} \label{eq:GD0_Im}
    \Im \Sigma^{{\rm sc}G D_0}_\nk(\veps)
    = -\pi \nint \frac{\dd\mb{q}}{V^{\rm BZ}} \sum_{m\nu} \, \abs{g_{mn\nu}(\mb{k},\mb{q})}^2
    \\
    \times \sum_{\pm} \bigl[f^\pm(\veps \pm \omega_\qnu) + n_\qnu \bigr]
    A^{{\rm sc}G D_0}_\mkq(\veps \pm \omega_\qnu) \,.
\end{multline}
The real part is obtained using the Kramers--Kronig relation, with the addition of the real-valued DW term:
\begin{equation} \label{eq:GD0_Re}
    \Re \Sigma^{{\rm sc}G D_0}_\nk(\veps)
    = \frac{1}{\pi} \mathcal{P} \! \nbint{-\infty}{\infty} \dd\veps' \frac{\Im \Sigma^{{\rm sc}G D_0}_\nk(\veps')}{\veps' - \veps}
    + \Sigma^{\rm DW}_\nk \,,
\end{equation}
where $\mathcal{P}$ denotes the principal value.
The spectral function in \Eq{eq:GD0_Im} is obtained via the Dyson equation:
\begin{align}
    \label{eq:GD0_A}
    A^{{\rm sc}G D_0}_\nk(\veps) &= -\frac{1}{\pi} \Im G^{{\rm sc}G D_0}_\nk(\veps) \,, \\
    \label{eq:GD0_G}
    G^{{\rm sc}G D_0}_\nk(\veps) &= \frac{1}{\veps - \veps_\nk + \mu - \Sigma^{{\rm sc}G D_0}_\nk(\veps)} \,.
\end{align}
Solving \Eqs{eq:GD0_Im}-\eqref{eq:GD0_G} self-consistently yields the sc$GD_0$ self-energy and spectral function.
We also iteratively adjust the temperature-dependent chemical potential to match the total electron occupation.

We remark that it is advantageous to first compute the imaginary part and then use the Kramers--Kronig relation to obtain the real part.
The relation between the real parts of $G$ and $\Sigma$ involves a $\bq$-dependent energy integral~\cite{Abramovitch2023}.
Separating \Eq{eq:GD0_Im} and \Eq{eq:GD0_Re} enables performing the $\bq$ integration for the imaginary part first, followed by a $\bq$-independent energy integral.

In practical \textit{ab initio} calculations, several hundred bands per atom are needed to converge the real part of the FM self-energy~\cite{Ponce2024}.
Calculating this via the Kramers--Kronig transformation is impractical, as it demands evaluating $\Im \Sigma(\veps)$ up to $\veps \sim 300~$eV.
We simplify the calculation by treating the contribution of high-energy bands as a static correction.
We thus self-consistently compute $\Im \Sigma^{{\rm sc}G D_0}_\nk(\veps)$ only for the states inside the active space window $[\veps^{\rm min}, \veps^{\rm max}]$, and add the static correction as
\begin{multline} \label{eq:GD0_Re_window}
    \Re \Sigma^{{\rm sc}G D_0}_\nk(\veps)
    = \frac{1}{\pi} \mathcal{P} \nbint{\veps^{\rm min}}{\veps^{\rm max}} \dd\veps' \frac{\Im \Sigma^{{\rm sc}G D_0}_\nk(\veps')}{\veps' - \veps}
    \\
    + \Re \Sigma^{G_0 D_0\tbar{\rm FM \tbar rest}}_\nk
    + \Re \Sigma^{\rm DW}_\nk \,.
\end{multline}
Here, $\Re \Sigma^{G_0 D_0\tbar{\rm FM \tbar rest}}_\nk$ is the one-shot FM self-energy [\Eq{eq:G0D0_FM}] from states outside the active space, $\veps_\mkq \notin [\veps^{\rm min}, \veps^{\rm max}]$, evaluated at the bare eigenvalue $\veps_\nk$.
This term can be efficiently computed by solving the linear Sternheimer equation~\cite{Gonze2011, Ponce2014} and interpolating using Wannier function perturbation theory (WFPT)~\cite{Lihm2021WFPT}.
The static corrections in the second line of \Eq{eq:GD0_Re_window} are always included during the self-consistent calculation of the sc$GD_0$ self-energy, i.e., \Eq{eq:GD0_Re_window} replaces \Eq{eq:GD0_Re}.
(See \Sec{supp-sec:active_space} of SM~\cite{\citeSupp} for additional details.)

In practice, we initialize the sc$GD_0$ self-energy as a constant $-i \eta$ with $\eta = 5$~meV.
We do not use any artificial broadening during the sc$GD_0$ iteration, and the choice of $\eta$ does not affect the converged sc$GD_0$ self-energy.
We iterate until the maximum change of $\Sigma_\nk^{{\rm sc}G D_0}(\veps)$ is below $0.1$~meV.
The sc$GD_0$ method is implemented as an extension of our \EPjl\ package~\cite{EPjl},
which uses quantities computed from \texttt{Quantum ESPRESSO}~\cite{Giannozzi2017},
Wannier90~\cite{Pizzi2020W90},
and EPW~\cite{Ponce2016EPW, Lee2023EPW}.
Computational details are provided in \Sec{supp-sec:comp_details} of SM~\cite{\citeSupp}.

The sc$GD_0$ method, also known as the self-consistent Migdal, self-consistent first-Born, or the noncrossing approximation, has been applied to the Holstein~\cite{Capone2003, Bauer2011, Sakkinen2015A, Sakkinen2015B, Rademaker2016SCBA, Esterlis2018SCBA, Dee2019SCBA, Dee2020SCBA, Mitric2022, Mitric2023} and Anderson--Holstein models~\cite{Chen2016SCBA}.
This nonperturbative method sums the infinite series of noncrossing rainbow diagrams [see \Fig{supp-fig:feynman}(c) of SM~\cite{\citeSupp}], including those with an arbitrary number of $e$-ph vertices.
It is expected to be valid from weak to intermediate coupling but will fail in the strong coupling regime, as in the case of self-trapped small polarons~\cite{Franchini2021Review}.
State-of-the-art \textit{ab initio} calculations~\cite{Abramovitch2023, Abramovitch2024} used a one-shot calculation of Eqs.~(\ref{eq:GD0_Im}, \ref{eq:GD0_Re}) (without $\Sigma^{\rm DW}_\nk$) with the spectral function obtained from dynamical mean-field theory to compute the $e$-ph self-energy.
Here, we self-consistently solve Eqs.~(\ref{eq:GD0_Im}-\ref{eq:GD0_Re_window}) to include nonperturbative $e$-ph effects on the spectral function.

We apply the sc$GD_0$ method to monolayer InSe, which has a Mexican-hat-like valance band~\cite{Zolyomi2014InSe, Rybkovskiy2014InSe, Debbichi2015InSe, Kibirev2018InSe}.
This dispersion leads to strong $e$-ph effects in hole-doped InSe monolayers: the one-shot $G_0 D_0$ spectral functions show a dynamic splitting of 150~meV or larger~\cite{Lugovskoi2019InSe}.
We study whether this large renormalization persists under a self-consistent update of the spectral function.

Figures~\ref{fig:spectral}(a)-(d) show the valence band spectral functions of monolayer InSe at 100~K, compared to the ARPES photointensity~\cite{Hamer2019InSe} measured at the same temperature.
The RS dispersion has several kinks with inverted curvature near the valence band maximum (VBM), an artifact also observed for the Fr\"ohlich model~\cite{Larsen1966}.
The cumulant spectral function shows a single broad peak with no satellites and a prominent dip at $\Gamma$.
The one-shot $G_0 D_0$ method yields two dynamically split bands with a splitting of 170~meV, consistent with Ref.~\cite{Lugovskoi2019InSe}.
However, this large splitting vanishes under self-consistency.
The sc$GD_0$ spectral function shows a broadened quasiparticle peak with multiple satellites, with a distance between the peaks of 30--50~meV, comparable to the longitudinal optical (LO) phonon energy of 27~meV.
An analysis of the difference between $G_0 D_0$ and sc$GD_0$ self-energies is provided in the End Matter (\Fig{fig:selfen}).
Our sc$GD_0$ result is consistent with the absence of a 150~meV scale gap in the ARPES photointensity~\cite{Kibirev2018InSe, Hamer2019InSe}.
Although both the sc$GD_0$ and cumulant spectral functions are consistent with the experimental observation within the experimental resolution, higher-resolution ARPES may observe the satellites predicted by our sc$GD_0$ calculation.


Beyond ARPES predictions that can sometimes be challenging to verify experimentally, the electron spectral function can be directly used to compute carrier conductivity, a simpler quantity to measure.
Neglecting the current vertex corrections, we calculate the conductivity in the bubble approximation, which is given by~\cite{MahanBook, Basov2011RMP}
\begin{multline} \label{eq:sigma_bubble}
    \Re \sigma^{\rm Bubble}_{\alpha\alpha}(\Omega)
    = \frac{\pi}{V^{\rm uc}} \sum_{mn} \nint \frac{\dd\mb{k}}{V^{\rm BZ}} \abs{v^\alpha_\mnk}^2  \\
    \times \nint \dd\veps \, \frac{f^+(\veps) - f^+(\veps + \Omega)}{\Omega}
    A_\mk(\veps) A_\nk(\veps + \Omega) \,.
\end{multline}
The corresponding mobility is $\mu_{\alpha\alpha}(\Omega) = \Re \sigma_{\alpha\alpha}(\Omega) / n^{\mathrm{c}}$ with $n^{\mathrm{c}}$ the carrier concentration.
\Equ{eq:sigma_bubble} has been used to calculate the conductivity due to electron-electron interaction~\cite{Basov2011RMP}, $e$-ph interaction~\cite{Zhou2019STO, Chang2022, Quan2024}, or both~\cite{Abramovitch2023, Abramovitch2024}.
Here, we investigate how the self-consistent renormalization of the spectral function affects the dc and ac mobility of monolayer InSe.

To study transport in the intrinsic limit, we set a small carrier concentration, placing the chemical potential within the band gap.
In this nondegenerate regime, mobility should be independent of carrier concentration~\cite{SzeBook, LundstromBook}.
However, as detailed in the End Matter, we find that both the $G_0 D_0$ and cumulant spectral functions yield strongly $n^{\mathrm{c}}$-dependent mobility (\Fig{fig:doping}).
We attribute this artifact to the slow $1/\veps^4$ decay of the spectral function above the band edge, which induces a large number of carriers on the tail of the spectral function.
To address this problem, we developed a truncation scheme that removes the tail above a cutoff energy (see \Sec{supp-sec:decay}of SM~\cite{\citeSupp}), which results in a $n^{\mathrm{c}}$-independent $G_0 D_0$ mobility.
However, this scheme cannot be applied to the cumulant spectral function of InSe due to its large broadening.
The sc$GD_0$ method does not suffer from this problem and yields an $n^{\mathrm{c}}$-independent mobility.

\begin{figure}[t]
\centering
\includegraphics[width=0.99\linewidth]{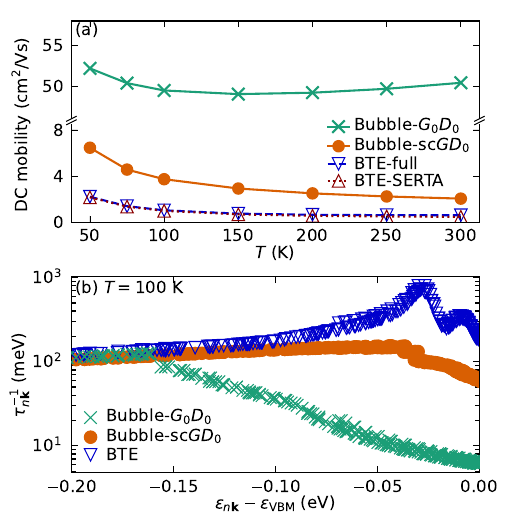}
\caption{
(a) Intrinsic dc mobility of monolayer InSe obtained within the bubble approximation with the one-shot (Bubble-$G_0 D_0$) or the self-consistent (Bubble-sc$GD_0$) spectral function, the full Boltzmann transport equation (BTE-full), and the BTE with the self-energy relaxation time approximation (BTE-SERTA).
(b) Effective scattering rates at $T=100$~K, calculated as the imaginary part of the self-energy evaluated at the bare or renormalized energy [Eqs.~(\ref{eq:inv_tau_BTE}-\ref{eq:inv_tau_GD0})].
}
\label{fig:InSe_dc}
\end{figure}

\Figu{fig:InSe_dc}(a) shows the intrinsic dc mobility of InSe.
While the BTE and self-energy relaxation time approximation (SERTA) mobilities~\cite{Ponce2018} are nearly identical and agree with the results from Refs.~\cite{Li2019InSe, Ponce2023PRB}, the bubble mobility with the sc$GD_0$ spectral function is 3 to 4 times larger.
This suggests that the quasiparticle approximation significantly underestimates the mobility.
Using the one-shot $G_0 D_0$ spectral function instead of the sc$GD_0$ one leads to an overestimation by more than an order of magnitude, and the mobility strongly depends on the artificial broadening $\eta$, varying inversely with $\eta$ [see \Fig{supp-fig:dc_sc_broadening} of SM~\cite{\citeSupp}].
These results can be understood in terms of the effective electron scattering rates shown in \Fig{fig:InSe_dc}(b), which are inversely proportional to the mobility.
The SERTA and BTE mobilities are linked to the the imaginary part of the $G_0 D_0$ self-energy at the bare band energy,
\begin{equation} \label{eq:inv_tau_BTE}
    (\tau^{\rm BTE}_\nk)^{-1} = 2 \abs{\Im \Sigma^{G_0 D_0}_\nk(\veps_\nk)} \,.
\end{equation}
For the spectral functions, we define the effective scattering rates using the self-energy at the renormalized quasiparticle energy,
\begin{align}
    \label{eq:inv_tau_G0D0}
    (\tau^{{\rm Bubble\tbar} G_0 D_0}_\nk)^{-1} &= 2 \abs{\Im \Sigma^{G_0 D_0}_\nk(E^{G_0 D_0}_\nk)} \,,
    \\
    \label{eq:inv_tau_GD0}
    (\tau^{{\rm Bubble\tbar} {\rm sc}G D_0}_\nk)^{-1} &= 2 \abs{\Im \Sigma^{{\rm sc}G D_0}_\nk(E^{{\rm sc}G D_0}_\nk)} \,.
\end{align}
The renormalized quasiparticle energies, $E^{G_0 D_0}_\nk$ and $E^{{\rm sc}G D_0}_\nk$, are the energies
where the spectral functions $A^{G_0 D_0}_\nk(\veps)$ and $A^{{\rm sc}G D_0}_\nk(\veps)$ are maximal.
Near the VBM, the BTE scattering rates exhibit a peak, while the sc$GD_0$ result shows a smoother variation and smaller values due to the self-consistent broadening.
For the one-shot self-energy, the scattering rates near the VBM are tiny because the renormalized valence bands lie within the bare band gap where the imaginary part of the $G_0 D_0$ self-energy is small.
This behavior is an artifact of the one-shot approximation.

\begin{figure}[tb]
\centering
\includegraphics[width=0.99\linewidth]{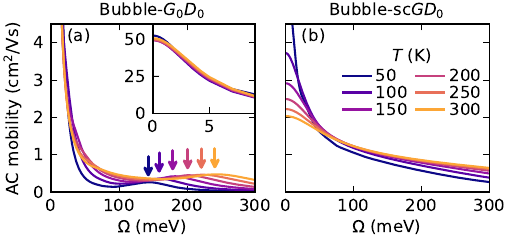}
\caption{
Intrinsic ac mobility of monolayer InSe computed using the bubble approximation~[\Eq{eq:sigma_bubble}] with the (a) $G_0 D_0$ and (b) sc$GD_0$ spectral functions.
The arrows indicate peaks at finite frequency.
The inset of panel (a) is an enlargement of the low-frequency Drude peak.
}
\label{fig:InSe_ac}
\end{figure}

Finally, we show in \Fig{fig:InSe_ac} the ac mobility calculated using the $G_0 D_0$ and sc$GD_0$ spectral functions.
For the one-shot spectral function, we observe a peak around 150--250~meV, corresponding to the gap shown in \Fig{fig:spectral}(c), as in Ref.~\cite{Lugovskoi2019InSe}.
This feature persists with a larger artificial broadening (see \Fig{supp-fig:ac_sc_broadening} of SM~\cite{\citeSupp}).
However, with the self-consistent spectral function, the unphysical peaks are smoothed out due to the reduced splitting and the broader Drude peak.

The self-consistent framework developed in this work offers a versatile tool for analyzing the nonperturbative effect of $e$-ph coupling with moderate computational effort, bridging the gap between one-shot approximations and numerically exact methods.
While we have focused on the spectral functions and conductivity, this framework can be extended to other properties such as superconductivity and indirect absorption~\cite{Noffsinger2012, Tiwari2024}.
Our work also lays the groundwork for applying diagrammatic many-body approaches to $e$-ph systems.
The sc$GD_0$ method could be extended to explore contributions beyond the Fan--Migdal diagram.
Notably, the development of a fully self-consistent $GD$ scheme using the renormalized phonon Green's function $D$~\cite{Sakkinen2015A, Sakkinen2015B, Berges2023},
the inclusion of the dynamical screening~\cite{Macheda2024, Lihm2024PlPh, Krsnik2024},
and the consideration of $e$-ph vertex correction~\cite{Esterlis2018SCBA, Dee2020SCBA, Lee2020TwoPh} is a promising avenue for future work.
Combining the sc$GD_0$ formalism with dynamical mean-field theory~\cite{Georges1996, Kotliar2006DMFT} and its diagrammatic extensions~\cite{Rohringer2018} to study transport from the interplay of electron-electron~\cite{Basov2011RMP} and $e$-ph coupling~\cite{Ponce2021} beyond the one-shot approximation~\cite{Abramovitch2023, Abramovitch2024} is another interesting direction for future research.

\begin{acknowledgments}
We thank Neil R.\ Wilson and Vladimir I.\ Fal'ko for kindly providing the data for Ref.~\cite{Hamer2019InSe},
and the anonymous referees who provided invaluable insights that led to an improved version of this work.
S.P. is a Research Associate of the Fonds de la Recherche Scientifique - FNRS.
This work was supported by the Fonds de la Recherche Scientifique - FNRS under Grants number T.0183.23 (PDR) and T.W011.23 (PDR-WEAVE).
This publication was supported by the Walloon Region in the strategic axe FRFS-WEL-T.
Computational resources have been provided by the EuroHPC JU award granting access to MareNostrum5 at Barcelona Supercomputing Center (BSC), Spain (Project ID: EHPC-EXT-2023E02-050), and by the Consortium des \'Equipements de Calcul Intensif (C\'ECI), funded by the FRS-FNRS under Grant No.~2.5020.11 and computational resources on Lucia, the Tier-1 supercomputer of the Walloon Region with infrastructure funded by the Walloon Region under the grant agreement n°1910247.

\textit{Data availability}---The data that support the findings of this article are openly available on the Materials Cloud Archive~\cite{MaterialsCloudArchive}.
\end{acknowledgments}

\FloatBarrier 

\makeatletter\@input{xy.tex}\makeatother
\bibliography{main}

\newpage

\onecolumngrid
\medskip
\begin{center}
    \textbf{\large End Matter}
\end{center}
\medskip
\twocolumngrid

\appendix


\begin{figure*}[tb]
\centering
\includegraphics[width=0.99\linewidth]{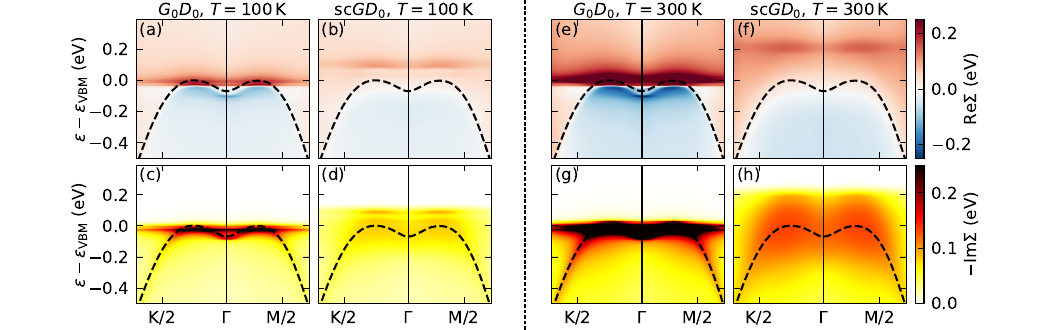}
\caption{
    (a, b) Real and (c, d) imaginary parts of the $G_0 D_0$ and sc$GD_0$ self-energies of monolayer InSe at $T=100$~K.
    (e-h) Same for $T=300$~K.
    The dashed black curves represent the bare electron band.
}
\label{fig:selfen}
\end{figure*}

\textit{Comparison of the self-energies}---%
\Figu{fig:selfen} compares the $G_0 D_0$ and sc$GD_0$ self-energies at 100~K and 300~K.
The one-shot $G_0 D_0$ self-energy shows a sharp peak in the imaginary part at $\veps = \veps_{\rm VBM} - \omega_{\rm LO}$, where $\omega_{\rm LO} \sim 27~\mathrm{meV}$ is the LO phonon frequency.
This peak, associated with phonon emission, is significantly damped and smeared in the sc$GD_0$ self-energy due to the self-consistent broadening of the spectral function.
This self-consistent regularization becomes more pronounced at higher temperatures.

At 300~K, the sc$GD_0$ self-energy reaches around 0.1~eV.
While this value is smaller than the full bandwidth of the valence band of InSe (1.2~eV; see \Fig{supp-fig:InSe_itp} of SM~\cite{\citeSupp}), it is larger than the width of the Mexican-hat feature at the valence band edge (0.07~eV).
Thus, the self-energy leads to a significant renormalization near the valence band edge, qualitatively altering the spectral function.

\begin{figure*}[htb]
\centering
\includegraphics[width=0.49\linewidth]{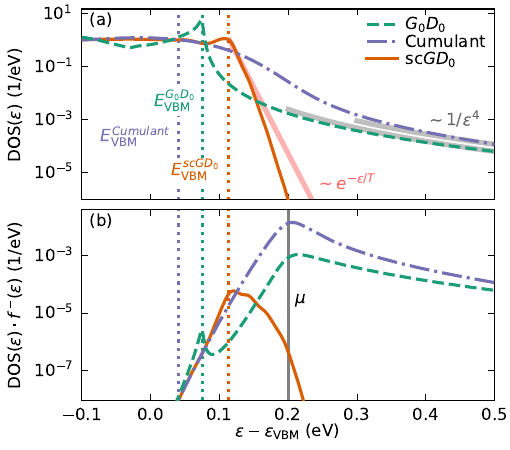}
\includegraphics[width=0.49\linewidth]{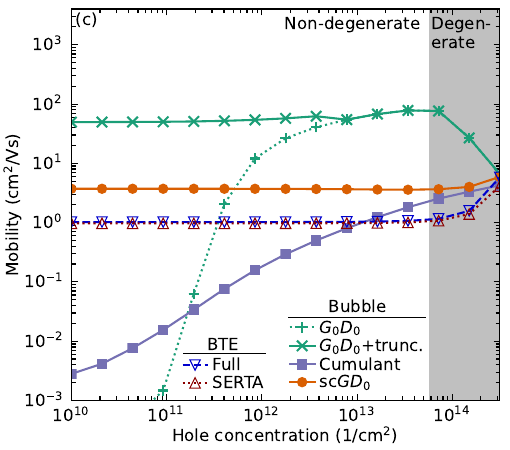}
\caption{
    (a) Density of states (DOS) and (b) hole DOS of monolayer InSe at $T=100$~K with a fixed chemical potential of 0.2~eV.
    The vertical dotted lines indicate the renormalized band edge, defined as the peak of the spectral function at the valence band maximum (VBM).
    The shaded curves in (a) show the power law and exponential decay derived in \Sec{supp-sec:decay} of SM~\cite{\citeSupp}.
    Both $G_0 D_0$ and cumulant spectral functions exhibit a slow $1/\veps^4$ decay in the midgap region, leading to an unphysical peak at $\mu$ in the hole DOS.
    In contrast, the sc$GD_0$ spectral function decays faster than $e^{-\veps / T}$, with the hole DOS peaking at the renormalized band edge as expected.
    (c) Doping dependence of the hole mobility of monolayer InSe at $T=100~$K.
    The white (gray) background indicates the nondegenerate (degenerate) semiconductor regime, where the BTE chemical potential is above (below) the bare VBM.
    Significant doping dependence is observed for the $G_0 D_0$ (green plus markers) and cumulant (blue squares) mobilities, while the mobility is doping-independent in the nondegenerate regime for the sc$GD_0$ (orange circles) and BTE results (blue and red triangles).
}
\label{fig:doping}
\end{figure*}

\textit{Tail of the spectral functions}---%
Compared to the $G_0 D_0$ and cumulant spectral functions, the sc$GD_0$ spectral functions show a clear qualitative difference in insulators and nondegenerate semiconductors: the former exhibits an unphysically long tail in the midgap region, which is absent in the latter.
Figures~\ref{fig:doping}(a, b) show the density of states (DOS)
\begin{equation}
    \mathrm{DOS}(\veps) = \sum_\nk A_\nk(\veps) \,,
\end{equation}
and the hole DOS ($f^{-}(\veps) \cdot \mathrm{DOS}(\veps)$).
In Fig.~\ref{fig:doping}(a), the DOS from the $G_0 D_0$ and the cumulant methods shows a slow $1 / \varepsilon^4$ decay (gray shaded curves).
This is a result of the Lorentzian broadening of the bare Green's function (see \Sec{supp-sec:decay} of SM~\cite{\citeSupp} for a detailed derivation.)
Importantly, reducing the artificial broadening $\eta$ reduces the amplitude of the tail but does not change the $1/\varepsilon^4$ scaling.
When multiplied by the hole occupation factor, which increases exponentially with energy up to the chemical potential, the hole DOS continues to increase above the renormalized band edge, peaking at the chemical potential [Fig.~\ref{fig:doping}(b)].
As a result, in nondegenerate semiconductors, the $G_0 D_0$ and cumulant methods incorrectly predict an unphysically large amount of carriers in the midgap region, where the spectral function is small.
In contrast, the sc$GD_0$ spectral function decays faster than $e^{-\varepsilon / T}$ (red shaded curve), and the hole DOS peaks at the renormalized band edge as expected.

\textit{Doping dependence of the conductivity}---%
Figure~\ref{fig:doping}(c) shows the doping dependence of the hole mobility of monolayer InSe at $100~\mathrm{K}$.
Surprisingly, the $G_0 D_0$ (green plus markers) and cumulant mobility (blue squares) show significant changes as a function of hole concentration.
This result contrasts sharply with what is typically observed in nondegenerate semiconductors with dominant $e$-ph scattering~\cite{SzeBook, LundstromBook}.
We attribute this artifact to the unphysical tail in the spectral functions, which places most of the holes near the chemical potential, where the actual spectral function is small.
Because of this strong doping dependence, the intrinsic (low doping limit) mobility cannot be properly calculated using the $G_0 D_0$ and cumulant approximations.
To alleviate this problem, we developed a truncation scheme [see \Eq{supp-eq:trunc_occ_p_doped} of SM~\cite{\citeSupp}] that removes the tail of the spectral function beyond the local minima in the hole DOS.
Then, we obtain a doping-independent Bubble-$G_0 D_0$ mobility (green crosses).
However, this truncation could not be applied to the cumulant spectral functions of InSe due to its large broadening.

In contrast, the sc$GD_0$ mobility (orange circles) as well as the BTE and SERTA mobility (blue and red triangles) remain doping-independent in the nondegenerate regime (white background).
Doping dependence is observed only in the metallic regime (gray background), where the chemical potential lies within the valence band.
Therefore, it is highly desirable to use the sc$GD_0$ method, rather than the $G_0 D_0$ or cumulant approximations, for studying transport in nondegenerate semiconductors.

\end{document}


\title{Supplemental Material:\\
Nonperturbative self-consistent electron-phonon spectral functions and transport}

\author{Jae-Mo Lihm\,\orcidlink{0000-0003-0900-0405}}
\email{jaemo.lihm@gmail.com}
\affiliation{%
European Theoretical Spectroscopy Facility and Institute of Condensed Matter and Nanosciences, Université catholique de Louvain, Chemin des Étoiles 8, B-1348 Louvain-la-Neuve, Belgium.
}%
\author{Samuel Ponc\'e\,\orcidlink{0000-0003-1159-8389}}
\email{samuel.ponce@uclouvain.be}
\affiliation{%
European Theoretical Spectroscopy Facility and Institute of Condensed Matter and Nanosciences, Université catholique de Louvain, Chemin des Étoiles 8, B-1348 Louvain-la-Neuve, Belgium.
}%
\affiliation{%
WEL Research Institute, avenue Pasteur, 6, 1300 Wavre, Belgium.
}%

\date{\today}

\maketitle

\section{\changed{Feynman diagrams for the self-energy approximations}} \label{sec:feynman}

\begin{figure}[htbp]
\centering
\includegraphics[width=0.99\linewidth]{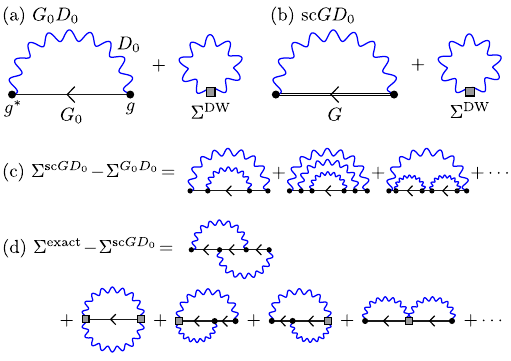}
\caption{
  Feynman diagrams for (a) the $G_0 D_0$ self-energy, (b) the sc$GD_0$ self-energy, (c) the difference between the two up to the sixth order in $g$
  \todo{Include DW}, and (d) the diagrams not included in the sc$GD_0$ method up to the fourth order.
  The solid lines represent the bare electron Green's function, the double line represents the renormalized electron Green's function, the wavy lines represent the phonon Green's function, and the black dots and gray squares represent the first- and second-order $e$-ph coupling vertex, respectively.
  The first line in (d) is the $e$-ph vertex correction, and the second line is due to the nonlinear $e$-ph coupling.
}
\label{fig:feynman}
\end{figure}

\Figu{fig:feynman} shows the Feynman diagrams for the $G_0 D_0$ self-energy, the sc$GD_0$ self-energy, and the difference between the two.
In the $G_0 D_0$ method, the self-energy is calculated up to the second order in the $e$-ph coupling $g$, and the bare electron Green's function enters the self-energy diagram.
In the sc$GD_0$ method, the renormalized electron Green's function is used instead.
The difference between the two methods lies in the inclusion of the higher order self-energy diagrams in which the phonon lines do not cross, as shown in \Fig{fig:feynman}(c).

\Figu{fig:feynman}(d) shows the diagrams not included in the sc$GD_0$ method.
The first line shows the $e$-ph vertex correction.
The cumulant method approximately includes this diagram~\cite{Dunn1975}.
The second line shows contributions due to the nonlinear $e$-ph coupling other than the simple Debye--Waller term~\cite{Nery2022}.

We note that the lowest-order Debye--Waller self-energy shown in \Figs{fig:feynman}(a, b) is static and real.
It is computed using the rigid-ion approximation~\cite{Allen1976, Allen1981, Allen1983, Gonze2011, Ponce2014PRB} and the momentum operator representation~\cite{Lihm2020}.

\section{\changed{Decay properties of the spectral functions}} \label{sec:decay}

\subsection{Derivation of the decay properties}

In the $G_0 D_0$ and cumulant approximations, the spectral function is calculated using the FM self-energy with an artificial broadening parameter $\eta$.
The bare Green's functions are broadened with a Lorentzian function which has a $1/\veps^2$ decay.
This slow decay leads to an unphysical tail in the spectral function, which is problematic for studying transport in lightly doped semiconductors.
Here, we analyze the decay of the spectral functions for insulators and lightly doped semiconductors.

For the $G_0 D_0$ case, for energy deep in the band gap $\veps_\mathrm{VBM} \ll \veps \ll \veps_\mathrm{CBM}$, but closer to the VBM than the CBM, the imaginary part of the self-energy becomes
\begin{align} \label{eq:tail_G0D0_selfen}
    &\Im \Sigma_\nk^{G_0 D_0}(\veps)
    \nnnl
    &\approx -\sum_{\nu m} \nint\! \frac{\dd\bq}{V^\mathrm{BZ}} \abs{g_{mn\nu}(\bk, \bq)}^2 \,
    ( 2n_{\bq\nu} + 1 ) \, \frac{\eta}{(\veps - \veps_\mkq)^2}
    \nnnl
    &\sim \frac{\eta}{(\veps - \veps_\mathrm{VBM})^2} \,.
\end{align}
The corresponding spectral function reads
\begin{align} \label{eq:tail_G0D0}
    A^{G_0 D_0}_\nk(\veps)
    &\approx -\frac{\Im \Sigma^{G_0 D_0}_\nk(\veps)}{\pi (\veps - \veps_\nk - \Re \Sigma^{G_0 D_0}_\nk(\veps))^2}
    \nnnl
    &\sim \frac{\eta}{(\veps - \veps_\mathrm{VBM})^2 (\veps - E^{G_0 D_0}_\mathrm{VBM})^2} \,,
\end{align}
where $E^{G_0 D_0}_\mathrm{VBM}$ is the renormalized VBM energy, defined as the peak position of the $G_0 D_0$ spectral function.
Thus, the spectral function decays as $1 / \veps^4$.
Importantly, reducing the broadening $\eta$ affects the amplitude of the tail, but does not change the $1/\veps^4$ scaling.

For the cumulant case, its spectral function can be written as a convolution in frequency space~\cite{Kandolf2022}:
\begin{equation} \label{eq:tail_Cum_conv}
  A^\mathrm{Cum}_\nk
  = A^\mathrm{QP}_\nk + A^\mathrm{QP}_\nk * A^\mathrm{S}_\nk
  + \frac{1}{2} A^\mathrm{QP}_\nk * A^\mathrm{S}_\nk * A^\mathrm{S}_\nk + \cdots \,.
\end{equation}
Here,
\begin{equation}
  A^\mathrm{QP}_\nk(\veps)
  = Z_\nk \, \delta \bigl(\veps - E^\mathrm{RS}_\nk \bigr) \,,
\end{equation}
is the quasiparticle spectral function, where
$E^\mathrm{RS}_\nk$ is defined in Eq.~(3) of the main text, and
\begin{equation}
  Z_\nk = \exp \biggl(\frac{\dd \Re \Sigma^{G_0 D_0}_\nk(\veps)}{\dd \veps} \Big\vert_{\veps = \veps_\nk} \biggr)
\end{equation}
is the quasiparticle weight.
$A^\mathrm{S}_\nk$ is the satellite term, defined as
\begin{equation}
  A^\mathrm{S}_\nk(\veps) = \frac{1}{\pi} \frac{\abs{\Im \Sigma_\nk^{G_0 D_0}(\veps + \veps_\nk)}}{\veps^2} \,.
\end{equation}
Due to the $1/\veps^2$ tail of the $G_0 D_0$ self-energy [\Eq{eq:tail_G0D0_selfen}], the satellite term decays as $1 / \veps^4$.
The decay of the first convolution term then reads
\begin{multline}
  \bigl( A^\mathrm{QP}_\nk * A^\mathrm{S}_\nk \bigr)(\veps)
  = \frac{Z_\nk}{\pi} \frac{\abs{\Im \Sigma_\nk^{G_0 D_0} \bigl(\veps - \Re \Sigma^{G_0 D_0}_\nk(\veps_\nk) \bigr)}}{\bigl( \veps - E^\mathrm{RS}_\nk \bigr)^2}
  \\
  \sim \frac{\eta}{
    \bigl(\veps - \veps_\mathrm{VBM} - \Re \Sigma^{G_0 D_0}_\nk(\veps_\nk) \bigr)^2
    \bigl( \veps - E^\mathrm{RS}_\nk \bigr)^2
  } \,.
\end{multline}
Higher order terms in the convolution in \Eq{eq:tail_Cum_conv} will decay faster.
Thus, the cumulant spectral function decays as $1 / \veps^4$:
\begin{equation} \label{eq:tail_cumulant}
    A^\mathrm{Cum}_\nk(\veps) \sim \frac{\eta}{
        \bigl(\veps - \veps_\mathrm{VBM} - \Re \Sigma^{G_0 D_0}_\nk(\veps_\nk) \bigr)^2
        \bigl( \veps - E^\mathrm{RS}_\nk \bigr)^2
    } \,.
\end{equation}

Now for the sc$GD_0$ case, we take the $\veps \gg \veps_\mathrm{VBM}$ limit of the sc$GD_0$ self-energy [Eq.~(8)] and find
\begin{multline} \label{eq:GD0_Im}
  \Im \Sigma^{\mathrm{sc}G D_0}_\nk(\veps)
  \approx -\pi \nint \frac{\dd\mb{q}}{V^\mathrm{BZ}} \sum_{m\nu} \, \abs{g_{mn\nu}(\mb{k},\mb{q})}^2
  \\
  \times n_\qnu \, A^{\mathrm{sc}G D_0}_\mkq(\veps - \omega_\qnu) \,.
\end{multline}
Here, we used the fact that since $A^{\mathrm{sc}G D_0}_\mkq(\veps)$ decays as $\veps$ increases, the $A^{\mathrm{sc}G D_0}_\mkq(\veps + \omega_\qnu)$ term (phonon emission) will be much smaller than the $A^{\mathrm{sc}G D_0}_\mkq(\veps - \omega_\qnu)$ contribution (phonon absorption).
Also, as we are considering the intrinsic limit (fully occupied valence band), we used $f^-(\veps) \approx 0$.
Assuming the same asymptotic behavior at all states, and approximating that all phonon frequencies are the same $\omega_\qnu \approx \omega_0$, we can write
\begin{equation} \label{eq:tail_scGD0_Sigma}
  \Im \Sigma^{\mathrm{sc}G D_0}(\veps)
  \approx -  \frac{\pi C^2}{e^{\omega_0 / T} - 1} \, A^{\mathrm{sc}G D_0}(\veps - \omega_0) \,,
\end{equation}
where $C$ is a constant factor with the dimension of energy.
From the Dyson equation, we find
\begin{equation} \label{eq:tail_scGD0_A}
  A^{\mathrm{sc}G D_0}(\veps)
  \approx - \frac{\Im \Sigma^{\mathrm{sc}G D_0}(\veps)}{\pi (\veps - \veps_\mathrm{VBM})^2} \,.
\end{equation}
Combining \Eqs{eq:tail_scGD0_Sigma} and \eqref{eq:tail_scGD0_A}, we find
\begin{equation}
  A^{\mathrm{sc}G D_0}(\veps)
  \approx \frac{C^2}{e^{\omega_0 / T} - 1} \, \frac{A^{\mathrm{sc}G D_0}(\veps - \omega_0)}{(\veps - \veps_\mathrm{VBM} - \omega_0)^2} \,.
\end{equation}
By taking logarithm of both sides, and assuming $\veps - \veps_\mathrm{VBM} \gg \omega_0$, one can convert this recursive relation into a differential equation
\begin{equation}
  \frac{\dd}{\dd \veps} \log A^{\mathrm{sc}G D_0}(\veps + \veps_\mathrm{VBM})
  = - \frac{1}{\omega_0} \Bigl( 2 \log (\veps) - \log \frac{c^2}{e^{\omega_0 / T} - 1} \Bigr) \,,
\end{equation}
whose solution is
\begin{multline} \label{eq:tail_scGD0_der}
  A^{\mathrm{sc}G D_0}(\veps + \veps_\mathrm{VBM})
  \sim \exp \Bigl[ - \frac{2 \veps}{\omega_0} \log \Bigl( \frac{\veps}{e C} \sqrt{e^{\omega_0 / T} - 1} \Bigr) \Bigr]
  \\
  \sim e^{-\veps / T}
  \exp \Bigl[ - \frac{2 \veps}{\omega_0} \log \Bigl( \frac{\veps}{e C} \sqrt{\frac{e^{\omega_0 / T} - 1}{e^{\omega_0 / T}}} \Bigr) \Bigr]
  \,.
\end{multline}
For $\veps$ large enough such that
\begin{equation}
  \frac{\veps}{e C}
  > \sqrt{\frac{\omega_0 + T}{\omega_0}}
  > \sqrt{\frac{e^{\omega_0 / T}}{e^{\omega_0 / T} - 1}} \,,
\end{equation}
the last exponent of \Eq{eq:tail_scGD0_der} is negative, and we find an exponential decay of the spectral function:
\begin{equation} \label{eq:tail_scGD0}
  A^{\mathrm{sc}G D_0}(\veps) < e^{-(\veps - \veps_\mathrm{VBM}) / T} \,.
\end{equation}

\subsection{Truncation scheme for the \texorpdfstring{$G_0 D_0$}{G0D0} and cumulant spectral functions}

As shown in \Fig{main-fig:doping} of the main text, the slow decay of the $G_0 D_0$ and cumulant spectral function leads to a strong doping dependence of the conductivity and makes it difficult to study transport in lightly doped systems.
Here, we propose a scheme to truncate the unphysical tail of the spectral function that alleviates this problem.

Let us consider a lightly hole-doped semiconductor in the intrinsic limit $\mu \gg \veps_\mathrm{VBM}$.
For the spectral tail \Eq{eq:tail_G0D0}, the hole occupation function reads
\begin{equation} \label{eq:trunc_occ}
    A^{G_0 D_0}_\nk(\veps) f^-(\veps)
    \sim \frac{\eta}{(\veps - \veps_\mathrm{VBM})^2 (\veps - E^{G_0 D_0}_\mathrm{VBM})^2} e^{(\veps - \mu) / T} \,.
\end{equation}
This function increases exponentially as the energy is increased.
Thus, one finds a large hole DOS at mid-gap energy [\Fig{main-fig:doping}(b)].
To solve this problem, we truncate the unphysical tail at some cutoff energy.
We use the minimum of the function in \Eq{eq:trunc_occ} as this cutoff, viewing the tail beyond this energy as unphysical.
The minimum is approximately located at
\begin{multline} \label{eq:trunc_e_p_doped}
    \veps^{{\rm trunc},\,p\tbar{\rm doped}}
    = \frac{\veps_\mathrm{VBM} + E^{G_0 D_0}_\mathrm{VBM}}{2}
    \\
    + 2T
    + \sqrt{4T^2 + \Bigl( \frac{\veps_\mathrm{VBM} - E^{G_0 D_0}_\mathrm{VBM}}{2} \Bigr)^2} \,,
\end{multline}
obtained from the analytical derivative of the approximated \Eq{eq:trunc_occ}.
Then, the truncated spectral function reads
\begin{equation} \label{eq:trunc_occ_p_doped}
    A^{G_0 D_0,\,p{\rm \tbar doped}}_\nk(\veps)
    = \Theta \bigl(\veps^{{\rm trunc},\,p\tbar{\rm doped}} - \veps \bigr) \, A^{G_0 D_0}_\nk(\veps) \,.
\end{equation}
For InSe at 100~K, the cutoff energy from the approximate minima [\Eq{eq:trunc_e_p_doped}] is $\veps^{{\rm trunc},\,p\tbar{\rm doped}} = 96$~meV, which is indeed close to the local minimum of the hole DOS $\veps = 89$~meV [\Fig{main-fig:doping}(b)].
Similarly, for lightly $n$-doped semiconductors, we truncate the spectral function at
\begin{multline}
    \veps^{{\rm trunc},\,n\tbar{\rm doped}}
    = \frac{\veps_{\rm CBM} + E^{G_0 D_0}_{\rm CBM}}{2}
    \\
    - 2T - \sqrt{4T^2 + \Bigl( \frac{\veps_{\rm CBM} - E^{G_0 D_0}_{\rm CBM}}{2} \Bigr)^2} \,,
\end{multline}
using the same sharp truncation
\begin{equation} \label{eq:trunc_occ_n_doped}
    A^{G_0 D_0,\,n{\rm \tbar doped}}_\nk(\veps)
    = \Theta \bigl(\veps - \veps^{{\rm trunc},\,n\tbar{\rm doped}} \bigr) \, A^{G_0 D_0}_\nk(\veps) \,.
\end{equation}
A similar cutoff can be derived for the cumulant approximation using \Eq{eq:tail_cumulant}.
We apply the truncation in \Eqs{eq:trunc_occ_p_doped} and \eqref{eq:trunc_occ_n_doped} only in the lightly-doped case, where $\mu > \veps^{{\rm trunc},\,p\tbar{\rm doped}}$ or $\mu < \veps^{{\rm trunc},\,n\tbar{\rm doped}}$, respectively.

This truncation scheme is applicable only when the truncation energy is well separated from the renormalized VBM or CBM.
Quantitatively, $\veps^{{\rm trunc},\,p\tbar{\rm doped}} - E^{G_0 D_0}_{\rm VBM} \approx 4T$, so the half-width of the spectral function should not exceed $4T$.
For InSe, this is the case for the $G_0 D_0$ spectral function but not for the cumulant one.
Hence, the local minimum in the hole DOS is found only in the former [\Fig{main-fig:doping}(b)], and we can apply the truncation scheme only to the $G_0 D_0$ method.
After truncation, we find a doping-independent mobility, as shown in \Fig{main-fig:doping}(c).

We emphasize that the sc$GD_0$ method does not have this problem with the tail, since its spectral function decays faster than the exponential factor $e^{-\veps / T}$ [\Eq{eq:tail_scGD0}].
This fact provides a strong reason to choose the sc$GD_0$ method over of the $G_0 D_0$ or the cumulant methods when studying transport in semiconductors.

\section{\changed{Active-space approximation of the self-energy}} \label{sec:active_space}

To avoid performing an infinite summation over bands to evaluate the self-energy, we use the active space approximation~\cite{Gonze2011, Ponce2014PRB}.
The active space $\mc{A} = \{ n\bk : \veps_\mathrm{min} \leq \veps_\nk \leq \veps_\mathrm{max} \}$ contains the all bands of interest.
For example, in transport calculations, the active space includes all the states with non-negligible carrier occupation.
To calculate the self-energy, we separate the sum over bands $m$ in the sc$GD_0$ self-energy [\Eq{main-eq:GD0_Im}] into two parts: one from states $\mkq$ in the active space, and one from the rest.
They read
\begin{widetext}
\begin{equation} \label{eq:sigma_active}
  \Im \Sigma^{\mathrm{sc}G D_0 \mathrm{\tbar FM\tbar active}}_\nk(\veps)
  = -\pi \nint \frac{\dd\mb{q}}{V^\mathrm{BZ}} \sum_{\nu}
  \!\!\!\! \sum_m^{\mkq \in \mc{A}} \!\!\!\!
  \, \abs{g_{mn\nu}(\mb{k},\mb{q})}^2
  \sum_{\pm} \bigl[f^\pm(\veps \pm \omega_\qnu) + n_\qnu \bigr] A^{\mathrm{sc}G D_0}_\mkq(\veps \pm \omega_\qnu) \,,
\end{equation}
\begin{equation} \label{eq:sigma_rest_der_1}
  \Im \Sigma^{\mathrm{sc}G D_0 \mathrm{\tbar FM\tbar rest}}_\nk(\veps)
  = -\pi \nint \frac{\dd\mb{q}}{V^\mathrm{BZ}} \sum_{\nu}
  \!\!\!\! \sum_m^{\mkq \notin \mc{A}} \!\!\!\!
  \, \abs{g_{mn\nu}(\mb{k},\mb{q})}^2
  \sum_{\pm} \bigl[f^\pm(\veps \pm \omega_\qnu) + n_\qnu \bigr] A^{\mathrm{sc}G D_0}_\mkq(\veps \pm \omega_\qnu) \,.
\end{equation}

We now make three approximations to \Eq{eq:sigma_rest_der_1}.
First, we neglect the renormalization of states $\mkq$ outside the active space and replace the spectral function with the delta function.
By also analytically performing the Kramers--Kronig transformation [\Eq{main-eq:GD0_Re}], we find
\begin{equation} \label{eq:sigma_rest_der_2}
  \Sigma^{\mathrm{sc}G D_0 \mathrm{\tbar FM \tbar rest}}_\nk(\veps)
  \approx
  \Sigma^{G_0 D_0 \mathrm{\tbar FM \tbar rest}}_\nk(\veps)
  = \nint \frac{\dd\mb{q}}{V^\mathrm{BZ}} \sum_{\nu}
  \!\!\!\! \sum_m^{\veps_\mkq \notin \mc{A}} \!\!\!\!
  \, \abs{g_{mn\nu}(\mb{k},\mb{q})}^2
  \sum_{\pm} \frac{f^\pm(\veps_\mkq) + n_\qnu}{\veps \pm \omega_\qnu - \veps_\mkq + i\eta} \,.
\end{equation}

Second, we use the static approximation $\veps \approx \veps_\nk$.
Since $\veps_\mkq$ (which is outside the active space window) is far from the $\veps_\nk$ (which is inside the active space), the integrand of \Eq{eq:sigma_rest_der_2} changes smoothly near $\veps \approx \veps_\nk$.
Although the static approximation is in principle valid only for $\veps$ near $\veps_\nk$, errors in the self-energy far from $\veps_\nk$ do not affect the spectral function significantly.
With these approximations, we find
\begin{equation} \label{eq:sigma_rest_der_3}
  \Sigma^{\mathrm{sc}G D_0 \mathrm{\tbar FM \tbar rest}}_\nk(\veps)
  \approx \Sigma^{G_0 D_0\tbar\mathrm{FM \tbar rest}}_\nk(\veps_\nk)
  = \nint \frac{\dd\mb{q}}{V^\mathrm{BZ}} \sum_{\nu}
  \!\!\!\! \sum_m^{\veps_\mkq \notin \mc{A}} \!\!\!\!
  \, \abs{g_{mn\nu}(\mb{k},\mb{q})}^2 \frac{f^\pm(\veps_\mkq) + n_\qnu}{\veps_\nk \pm \omega_\qnu - \veps_\mkq + i\eta} \,.
\end{equation}

Finally, we use the adiabatic approximation of neglecting the phonon frequency in the denominator.
This step is justified because $\abs{\veps_\nk - \veps_\mkq}$ is much larger than $\omega_\qnu$.
This approximation is commonly used in the calculation of the phonon-induced band structure renormalization~\cite{Gonze2011, Ponce2014PRB, Ponce2015, Ponce2024}.
Then, we arrive at our final expression
\begin{equation} \label{eq:sigma_rest_der_4}
  \Sigma^{\mathrm{sc}G D_0 \mathrm{\tbar FM \tbar rest}}_\nk(\veps)
  \approx \Sigma^{G_0 D_0\tbar\mathrm{FM \tbar rest}}_\nk(\veps_\nk)
  \approx \nint \frac{\dd\mb{q}}{V^\mathrm{BZ}} \sum_{\nu}
  \!\!\!\! \sum_m^{\veps_\mkq \notin \mc{A}} \!\!\!\!
  \, \abs{g_{mn\nu}(\mb{k},\mb{q})}^2 \frac{2n_\qnu + 1}{\veps_\nk - \veps_\mkq} \,.
\end{equation}
\end{widetext}
We use \Eq{eq:sigma_rest_der_4} to approximate the rest-space contribution to the $G_0D_0$ and sc$GD_0$ self-energies, as well as the cumulant spectral function.
This term is the rest-space contribution to the static on-shell self-energy~\cite{Ponce2015, Ponce2024}, which can also be Wannier interpolated using Wannier function perturbation theory~\cite{Lihm2021WFPT}.

\begin{figure}[tb]
\centering
\includegraphics[width=0.99\linewidth]{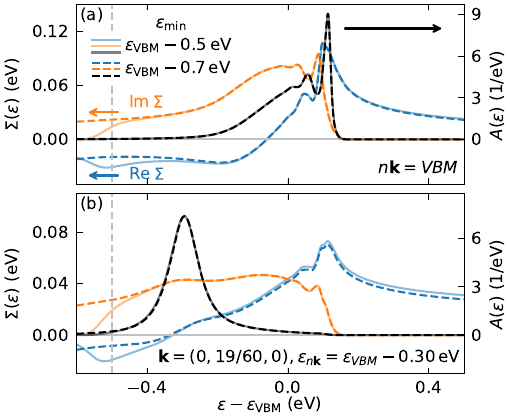}
\caption{
    Frequency dependence of the sc$GD_0$ self-energy and spectral function of monolayer InSe at 100~K at (a) the valence band maximum (VBM), and (b) $\mb{k} = (0, 19/60, 0)$ (in crystal coordinates), computed with a narrow (solid curves) and a wide (dashed curves) energy window.
    The self-energy from the two windows agree well at $\veps - \veps_\mathrm{VBM} > - 0.3~$eV.
    The spectral functions show a better agreement without any noticeable difference in the full energy range since its magnitude is small in the region where the self-energy discrepancy is larger.
}
\label{fig:InSe_selfen_window}
\end{figure}

\Figu{fig:InSe_selfen_window} compares the self-energy and spectral functions computed with two different energy windows, a narrower one ($\veps_\mathrm{min} - \veps_\mathrm{VBM} = -0.5~$eV, solid curves), and a wider one ($\veps_\mathrm{min} - \veps_\mathrm{VBM} = -0.7~$eV, dashed curves).
We find that right at the boundary of the narrower energy window ($\veps - \veps_\mathrm{VBM} = -0.5~$eV), the two self-energies are different.
Also, in \Fig{fig:InSe_selfen_window}(b), the self-energy at higher energies $\veps - \veps_\mathrm{VBM} > +0.2~$eV shows a small error of $\sim0.003~$eV ($\sim$10\% of the total self-energy), because we approximate $\veps \approx \veps_\nk$.
Still, the corresponding spectral functions, which are the physical quantities of interest, show a good agreement without any noticeable difference in the full energy range.
This is because the static approximation is accurate near $\veps_\nk$, where the spectral function is large, and errors are made at higher or lower energies, where the spectral function is small.

\section{Computational details} \label{sec:comp_details}

We performed DFT and DFPT calculations using \textsc{Quantum ESPRESSO}~\cite{Giannozzi2017}, with a 160~Ry kinetic energy cutoff and a 16$\times$16$\times$1 unshifted $\mb{k}$-point grid.
We used norm-conserving fully-relativistic pseudopotentials~\cite{Hamann2013ONCVPSP} in the Perdew-Burke-Ernzerhof (PBE) exchange-correlation functional~\cite{Perdew1996} from \textsc{PseudoDojo} (v0.4)~\cite{Setten2018}.
Spin-orbit coupling was not included.
We relaxed the lattice constant and atomic positions.
The relaxed in-plane lattice constant (4.086~\AA) agrees well with previous calculations~\cite{Zolyomi2014InSe, Debbichi2015InSe, Kibirev2018InSe} and the experimental bulk value (4.002~\AA).
We computed phonons for the undoped system and neglected phonon softening due to doping, which was included in Ref.~\cite{Lugovskoi2019InSe}.
A truncated Coulomb interaction for two-dimensional systems was applied in the DFT and DFPT calculations following Ref.~\cite{Sohier2017}.

\begin{table}[]
\centering
\begin{tabular}{crrrr}
    Atom
    & \multicolumn{1}{c}{$Z^*_{\parallel}$}
    & \multicolumn{1}{c}{$Z^*_{\perp}$}
    & \multicolumn{1}{c}{$Q_{xyx}$}
    & \multicolumn{1}{c}{$Q_{xzx}$} \\ \hline 
    In   & $2.444$  & $0.169$   & $-7.137$  & $  4.952$ \\  
    Se   & $-2.444$ & $-0.169$  & $-1.601$  & $-11.747$ \\  
\end{tabular}
\caption{Non-equivalent elements of the Born effective charge and quadrupole tensor of monolayer InSe in $e$ and $e\cdot\mathrm{Bohr}$ units.
Other nonzero elements follow from $Q_{xxy} = Q_{xyx} = -Q_{yyy}$ and $Q_{xzx} = Q_{yzy}$.
The out-of-plane components $Q_{\alpha\beta z}$ are not used.
}
\label{tab:quad}
\end{table}

\begin{figure}[tb]
\centering
\includegraphics[width=0.99\linewidth]{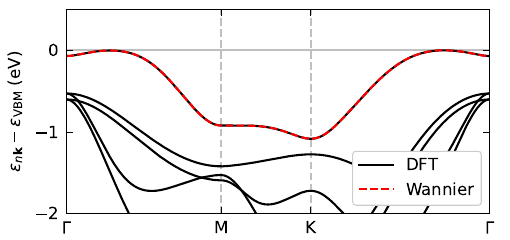}
\caption{
Band structure of monolayer InSe computed using DFT and Wannier interpolation.
}
\label{fig:InSe_bands}
\end{figure}

\begin{figure}[tb]
\centering
\includegraphics[width=0.99\linewidth]{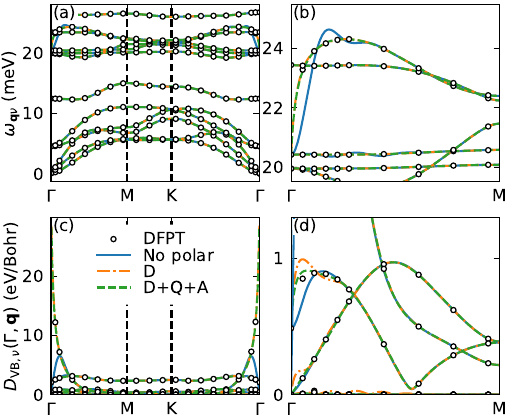}
\caption{
(a) Phonon frequency calculated using DFPT (white circles) and interpolated using Wannier--Fourier interpolation without any long-range treatment (No polar), using the two-dimensional dipole term (D) and using the dipole, quadrupole, and Berry connection terms (D+Q+A).
(b) Zoom of (a) along the $\Gamma$M direction.
(c, d) Same as (a, b) for the deformation potential.
}
\label{fig:InSe_itp}
\end{figure}

\begin{figure}[tb]
\centering
\includegraphics[width=0.99\linewidth]{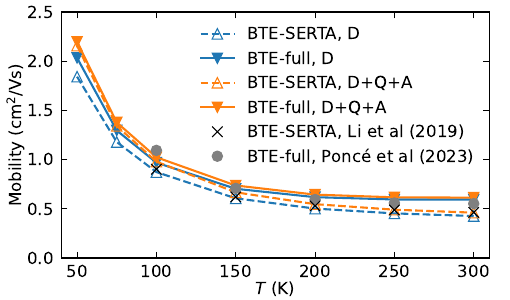}
\caption{
Comparison of the mobility of monolayer InSe computed using the long-range interatomic force constants and $e$-ph coupling computed using only the two-dimensional dipole term (D) and the dipole, quadrupole, and Berry connection terms (D+Q+A), compared with the calculation of Ref.~\cite{Li2019InSe} using a three-dimensional dipole term, and that of Ref.~\cite{Ponce2023PRB} using D+Q+A.
}
\label{fig:InSe_comparison}
\end{figure}

We used \textsc{Wannier90}~\cite{Pizzi2020W90} and EPW~\cite{Giustino2007EPW, Ponce2016EPW, Lee2023EPW} to construct localized Wannier functions and real-space matrix elements, sampling the Brillouin zone with a $16 \times 16 \times 1$ grid for both electrons and phonons.
We generated a single Wannier function using an $s$ orbital at the midpoint of two In atoms as an initial guess and performed maximal localization~\cite{Marzari1997}.
The interpolated band structure is shown in \Fig{fig:InSe_bands}.
We included the dipolar~\cite{Verdi2015, Sjakste2015}, quadrupolar~\cite{Brunin2020PRL, Brunin2020PRB, Jhalani2020, Park2020}, and Berry connection~\cite{Ponce2023PRL, Ponce2023PRB} contributions to the long-range intertomic force constants and $e$-ph coupling.
We modeled the two-dimensional electrostatic long-range interaction following Ref.~\cite{Royo2021}.
We computed the quadrupole tensors using the \textsc{Abinit} code~\cite{Royo2019, Gonze2020Abinit}, using pseudopotentials without nonlinear core correction due to the requirement of the implementation.
The in-plane dielectric constant was $\epsilon_\parallel = 3.772$.
Nonzero elements of the Born effective charges and quadrupole tensors are reported in Table~\ref{tab:quad}.

\begin{figure}[tb]
\centering
\includegraphics[width=0.99\linewidth]{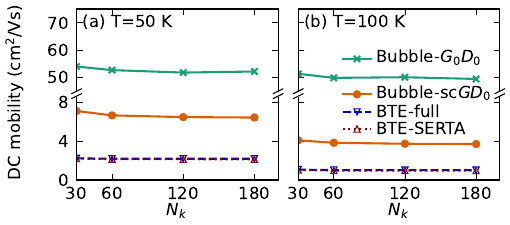}
\caption{
Convergence of the mobility of monolayer InSe at 50~K and 100~K with respect to the fine grid of the Brillouin zone sampling, where the $\bk$- and $\bq$-grids are both $N_k \times N_k \times 1$.
}
\label{fig:InSe_convergence}
\end{figure}

We performed Wannier interpolation and computed the sc$GD_0$ self-energy using an in-house developed code \EPjl\ written in the Julia programming language~\cite{Bezanson2017Julia}.
We used a $180 \times 180 \times 1$ sampling of $\mb{k}$ and $\mb{q}$ points,
filtered states with energy $\veps_\nk \in [\veps^{\rm min}, \veps^{\rm max}] = [-0.5, 0]~\mathrm{eV}$ with respect to the VBM energy,
and used a frequency grid of range [$-0.6$, 0.5]~eV with a spacing of 1~meV.
For the spectral function plots, we used a wider window including states with energy $\veps_\nk \in [\veps^{\rm min}, \veps^{\rm max}] = [-0.7, 0]~\mathrm{eV}$ and used a frequency grid of range [$-0.8$, 0.5]~eV with a spacing of 1~meV.
We used a 5~meV broadening for the $G_0 D_0$ and cumulant calculations, while we did not set any artificial broadening for the sc$GD_0$ calculation.
We numerically verified that the frequency integral sum rule of the spectral function at each state is satisfied closer than $1\pm 0.005$.

\Figu{fig:InSe_bands} shows that although the Wannier interpolated band is separated from the excluded bands at each $\bk$ point, their respective density of states overlaps.
Thus, a phonon mode with finite $\bq$ can resonantly couple the Wannierized and excluded bands.
This case was not considered in the original version of WFPT~\cite{Lihm2021WFPT}, where one assumed an energy window that separates Wannierized and excluded bands.
To deal with this issue, we add a $\eta^{\rm WFPT} = 100~\mathrm{meV}$ broadening for the $G_0 D_0$-FM self-energy~[\Eq{main-eq:G0D0_FM}] coming from the transition between the Wannierized and excluded valence bands.
In practice, $i \eta^{\rm WFPT}$ is added to the denominator of Eq.~(9) of Ref.~\cite{Lihm2021WFPT}.
This addition has only a small effect on the self-energy near the VBM, which is well separated from the excluded bands by a few hundred meV, while significantly improving the convergence of WFPT interpolation.

\Figu{fig:InSe_itp} compares the interpolation of phonon frequency and $e$-ph coupling using different long-range treatments.
For the phonon frequency, we find that the inclusion of the dipole term is sufficient to obtain an optical mode with a finite slope at $\Gamma$~\cite{Sohier2017}.
For the $e$-ph coupling, we show the deformation potential
\begin{equation} \label{eq:deformation_potential}
    D_{mn\nu}(\bk, \bq) = \Bigl[ 2 M \omega_\qnu \abs{g_{mn\nu}(\bk, \bq)}^2 \Bigr]^{1/2}
\end{equation}
for $m$ and $n$ corresponding to the highest valence band and $\bk = \mb{\Gamma}$.
Here, $M$ is the total mass of the unit cell.
Including the quadrupole and Berry connection terms~\cite{Ponce2023PRB, Ponce2023PRL} improves the interpolation near $\Gamma$.

\Figu{fig:InSe_comparison} compares the conductivity computed with different long-range treatments.
The effect of quadrupole and Berry connection terms on mobility is relatively minor.
We find a rough agreement with the mobility computed by \citet{Li2019InSe} using a three-dimensional dipole term~\cite{Verdi2015,Sjakste2015}, and that computed by \citet{Ponce2023PRB} using the dipole, quadrupole, and Berry connection terms.
We considered hole doping of $n^\mathrm{c}=-10^{9}~\mathrm{cm}^{-2}$, which is in the intrinsic limit.
\Figu{fig:InSe_convergence} shows the convergence of the mobility with respect to the $\mb{k}$ and $\mb{q}$ point sampling.
We find convergence at $N_k \geq 60$.

\clearpage


\onecolumngrid

\section{Additional results for monolayer InSe} \label{sec:InSe_additional}

\begin{figure*}[tb]
\centering
\includegraphics[width=0.99\linewidth]{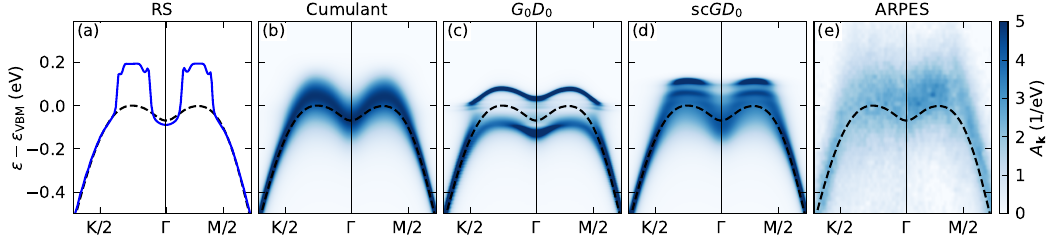}
\caption{
Same as \Fig{main-fig:spectral} of the main text, but without the Gaussian filter applied on the calculated spectral functions (b-d).
}
\label{fig:spectral_noblur}
\end{figure*}

\twocolumngrid

\begin{figure}[b]
\centering
\includegraphics[width=0.99\linewidth]{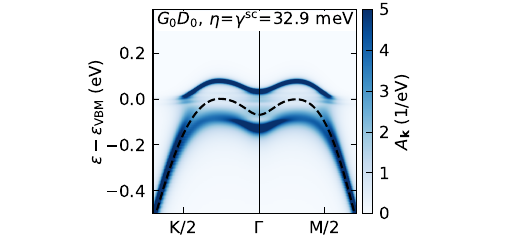}
\caption{
The $G_0 D_0$ spectral function of monolayer InSe at $T=100~\mathrm{K}$ as \Fig{fig:spectral_noblur}(c), but computed with the broadening parameter $\eta = \gamma^{\rm sc}(T=100~\mathrm{K}) = 32.9~\mathrm{meV}$ determined from the sc$GD_0$ self-energy~[\Eq{eq:eta_from_GD0}].
}
\label{fig:spectral_sc_broadening}
\end{figure}

\begin{figure}[h]
\centering
\includegraphics[width=0.99\linewidth]{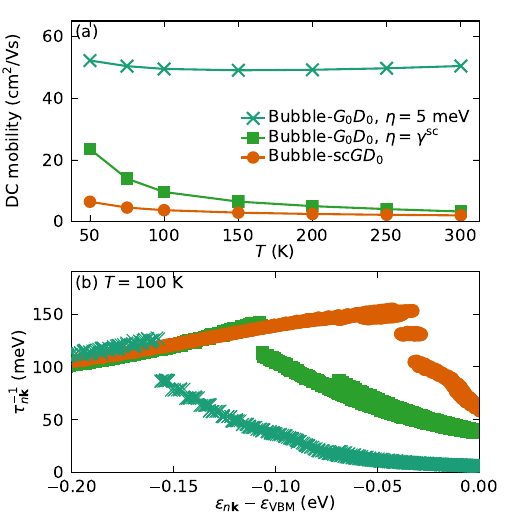}
\caption{
(a) Intrinsic DC mobility of monolayer InSe as in \Fig{main-fig:InSe_dc}(a) of the main text, but for the $G_0 D_0$ method with the broadening determined from the sc$GD_0$ self-energy~[\Eq{eq:eta_from_GD0}].
The self-consistent broadening is 12.9, 22.2, 32.9, 49.5, 61.4, 72.7, and 116.2~meV for $T=50$, 75, 100, 150, 200, 250, and 300~K, respectively.
(b) Effective inverse lifetimes at $T=100$~K as in \Fig{main-fig:InSe_dc}(b) of the main text.
}
\label{fig:dc_sc_broadening}
\end{figure}

\Figu{fig:spectral_noblur} shows the same calculated spectral functions as in \Fig{main-fig:spectral} of the main text, but without the Gaussian broadening filter applied.
The results are similar, but multiple satellites are clearly visible in the sc$GD_0$ spectral function \Fig{fig:spectral_noblur}(d).

The one-shot $G_0 D_0$ self-energy and spectral function depend sensitively on the value of the broadening parameter $\eta$ in the denominator of \Eq{main-eq:G0D0_FM}.
The two standard methods to determine the value of $\eta$ are (i) choosing the smallest possible $\eta$ that yields converged results for the size of the used $\bk$- and $\bq$-point mesh, and (ii) using the physical linewidth of the states~\cite{Lihm2024Piezo}.
We note that Ref.~\cite{Lugovskoi2019InSe} used an auxiliary broadening of $\eta \sim T$ to calculate the $G_0 D_0$ self-energy [\Eq{main-eq:G0D0_FM}].
In the second approach, the broadening parameter can be determined using the sc$GD_0$ calculation as a reference:
\begin{equation} \label{eq:eta_from_GD0}
    \eta = \gamma^{\rm sc}(T) = \abs{\Im \Sigma^{G D_0}_{\rm VBM}(E^{G D_0}_{\rm VBM},\, T)}\,.
\end{equation}

\begin{figure}[t]
\centering
\includegraphics[width=0.99\linewidth]{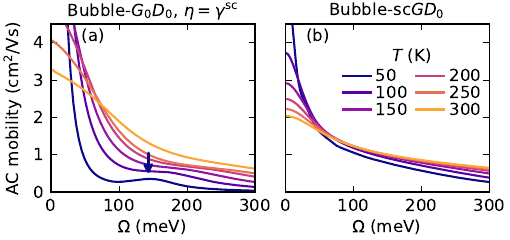}
\caption{
Intrinsic AC mobility of monolayer InSe as in \Fig{main-fig:InSe_ac}, but for the $G_0 D_0$ method with the broadening parameter determined from the sc$GD_0$ self-energy~[\Eq{eq:eta_from_GD0}].
}
\label{fig:ac_sc_broadening}
\end{figure}

In the main text, we used the first approach (with $\eta = 5~\mathrm{meV}$) and demonstrated that the $G_0 D_0$ spectral function differs significantly from the sc$GD_0$ one.
\Figu{fig:spectral_sc_broadening} shows that choosing $\eta$ using the calculated sc$GD_0$ linewidth still does not change the fact that a large gap appears.
Figures~\ref{fig:dc_sc_broadening} and \ref{fig:ac_sc_broadening} compare the $G_0 D_0$ mobility using the self-consistent broadening parameter with the sc$GD_0$ mobility.
Even though the broadening parameter is chosen to make the $G_0 D_0$ self-energy as similar as possible to the sc$GD_0$ result, we still find sizable qualitative differences between the mobilities.
Therefore, the full self-consistency of the frequency- and momentum-dependent spectral function is crucial.

\FloatBarrier  
\makeatletter\@input{xx.tex}\makeatother
\bibliography{main}